\documentstyle [12pt] {article}

\catcode`@=12
\begin{document}
\thispagestyle{empty}
\begin{flushright} UCRHEP-T261\\August 1999\
\end{flushright}
\vskip 0.5in
\begin{center}
{\Large \bf Hierarchical Four-Neutrino Oscillations\\
With a Decay Option\\}
\vskip 2.0in
{\bf Ernest Ma$^1$, G. Rajasekaran$^2$, Ion Stancu$^1$\\}
\vskip 0.3in
{$^1$ \sl Physics Department, Univeristy of California, Riverside, 
CA 92521, USA\\}
\vskip 0.1in
{$^2$ \sl Institute of Mathematical Sciences, Madras 600113, India\\}
\end{center}
\vskip 1.8in
\begin{abstract}\
We present a new and novel synthesis of all existing neutrino data regarding 
the disappearance and appearance of $\nu_e$ and $\nu_\mu$.  We assume four 
neutrinos: $\nu_e, \nu_\mu, \nu_\tau$, as well as a heavier singlet neutrino 
$\nu_s$ of a few eV.  The latter may decay into a massless Goldstone boson 
(the singlet Majoron) and a linear combination of the doublet antineutrinos.  
We comment on how this scenario may be verified or falsified in future 
experiments.
\end{abstract}
\newpage
\baselineskip 24pt

Accepting the totality of present experimental evidence for neutrino 
oscillations\cite{1,2,3}, it is not unreasonable to entertain the idea 
that there are four light neutrinos.  Since the invisible decay of the $Z$ 
boson tells us that there are only three light doublet neutrinos, i.e. 
$\nu_e, \nu_\mu, \nu_\tau$, the fourth light neutrino $\nu_s$ should be a 
singlet.  Usually, $\nu_s$ is assumed to mix with the other neutrinos in a 
$4 \times 4$ mass matrix for a phenomenological understanding\cite{4} of all 
the data.  However, given that $\nu_s$ is different from $\nu_{e,\mu,\tau}$, 
it may have some additional unusual property, such as decay.  In fact, as 
shown below, this is a natural consequence of the spontaneous breakdown 
of lepton number in the simplest model\cite{5}, and it has some very 
interesting and verifiable predictions in future neutrino experiments.

If only atmospheric\cite{1} and solar\cite{2} neutrino data are considered, 
then hierarchical three-neutrino oscillations with
\begin{eqnarray}
\nu_1 &=& \nu_e \cos \theta - {1 \over \sqrt 2} (\nu_\mu + \nu_\tau) 
\sin \theta, \\ \nu_2 &=& \nu_e \sin \theta + {1 \over \sqrt 2} (\nu_\mu + 
\nu_\tau) \cos \theta, \\ \nu_3 &=& {1 \over \sqrt 2} (-\nu_\mu + \nu_\tau),
\end{eqnarray}
where $m_1 << m_2 << m_3$, would fit the data very well.  Here $m_3^2 \sim 
10^{-3}$ eV$^2$, $(\sin^2 2 \theta)_{atm} = 1$, and $m_2^2 \sim 10^{-5}$ 
eV$^2$ for the matter-enhanced oscillation solution\cite{6} to the solar 
neutrino deficit with $(\sin^2 2 \theta)_{sol} \sim 10^{-3}$ or near 1, 
or $m_2^2 \sim 10^{-10}$ eV$^2$ for the vacuum oscillation solution with 
$(\sin^2 2 \theta)_{sol} \sim 1$.

We now add a fourth neutrino $\nu_s$ and assume that it mixes a little with 
$\nu_e$ and $\nu_\mu$ to explain the LSND data\cite{3}.  Since the relevant 
$\Delta m^2$ is now about 1 eV$^2$, it is natural to take $m_4^2 \sim 1$ 
eV$^2$, but this hierarchical solution is disfavored\cite{7}, because the 
observed $\bar \nu_\mu \to \bar \nu_e$ probability\cite{3} is contradicted by 
the $\nu_\mu \to \nu_\mu$ data of CDHSW\cite{8} together with the $\bar \nu_e 
\to \bar \nu_e$ data of Bugey\cite{9}.  However, there are two ways that 
this conclusion may be evaded.  (1) Let $m_4^2 \sim 25$ eV$^2$, then the 
constraint due to the CDHSW experiment is not a factor, but now there 
are three other accelerator $\nu_\mu \to \nu_e$ experiments: 
BNL-E734\cite{10}, BNL-E776\cite{11}, and CCFR\cite{12}, 
which have bounds close to but allowed by the LSND 99\% likelihood contour. 
This is a marginal hierarchical four-neutrino oscillation solution to all 
the data.  (2) If $\nu_4$ decays, then the parameter space for an acceptable 
solution should open up.  For example, in the CDHSW experiment, two detectors 
at different distances compare their respective $\nu_\mu$ fluxes and the 
ratio is taken.  If the $\nu_4$ component of $\nu_\mu$ decays away already 
before reaching the first detector, the ratio remains at unity.  In contrast 
to the case of only oscillations, this experiment is then unable to restrict 
$m_4^2$.  Not only that, since the argument\cite{7} against the hierarchical 
four-neutrino spectrum depends crucially on the CDHSW experiment, it is clear 
that it cannot be valid in general.

The idea of neutrino decay is of course not new.  It is naturally related to 
the spontaneous breakdown of lepton number\cite{5,13}.  The associated 
massless Nambu-Goldstone boson\cite{14} is called the Majoron and the 
typical decay $\nu_2 \to \bar \nu_1 +$ Majoron occurs if kinematically 
allowed.  The triplet Majoron\cite{13} is ruled out experimentally because 
the decay $Z \to$ Majoron + partner (imaginary and real parts respectively of 
the lepton-number carrying scalar field) would have counted as the equivalent 
of two extra neutrino flavors.  The singlet Majoron\cite{5} is unconstrained 
because it has no gauge interactions.  We assign lepton number $L = -1$ to 
$\nu_s$ and assume the existence of a scalar particle $\chi^0$ with 
$L = 2$.  [By convention, $\nu_s$ is left-handed.  If we use a right-handed 
singlet neutrino $\nu_R$ instead, then it would be assigned $L = +1$.] 
Hence the relevant terms of the interaction Lagrangian are given by
\begin{equation}
{\cal L}_{int} = g_s \nu_s \nu_s \chi^0 + \sum_{\alpha=e,\mu,\tau} h_\alpha 
\nu_s (\nu_\alpha \phi^0 - l_\alpha \phi^+) + h.c.
\end{equation}
As $\langle \chi^0 \rangle$ and $\langle \phi^0 \rangle$ become nonzero, 
$\nu_s$ becomes massive and also mixes with $\nu_{e,\mu,\tau}$ to form the 
mass eigenstates $\nu_{1,2,3,4}$.  At the same time, $\sqrt 2 Im \chi^0$ 
becomes the massless Majoron $M$ and the decay
\begin{equation}
\nu_4 \to \bar \nu_{1,2,3} + M
\end{equation}
is now possible.  Neutrino decay involving only $\nu_{e,\mu,\tau}$ was 
recently proposed\cite{15} to explain the atmospheric data\cite{1}, but 
that becomes a poor fit after the inclusion of the upward going 
muons\cite{16}.  More recently, it was shown\cite{17} that combining 
oscillation and decay (at the expense of also adding $\nu_s$) gives again a 
good fit.  In contrast, the effects we envisage here of $\nu_4$ decay in 
atmospheric and solar neutrino data are both small and do not change the usual 
oscillation interpretation appreciably, as shown below.

Let $\nu_{e,\mu,\tau,s}$ be related to the mass eigenstates $m_{1,2,3,4}$ 
through the unitary matrix $U_{\alpha i}$, which will be assumed real in the 
following for simplicity.  Let $m_4 >> m_3 >> m_2 >> m_1$ with $\nu_4$ having 
the decay lifetime $\tau_4$.  Then for solar and atmospheric neutrino 
oscillations with $m_4^2 L/4E >> 1$, the probability of $\nu_\alpha \to 
\nu_\beta$ is given by
\begin{equation}
P_{\alpha \beta} = \delta_{\alpha \beta} ( 1 - 2 U_{\alpha 4}^2 ) + 
U_{\alpha 4}^2 U_{\beta 4}^2 ( 1 + x^2 ) - 4 \sum_{i<j<4} 
U_{\alpha i} U_{\alpha j} U_{\beta i} U_{\beta j} \sin^2 {\Delta m_{ij}^2 L 
\over 4 E},
\end{equation}
where
\begin{equation}
x = e^{-m_4 L/2 E \tau_4}.
\end{equation}
In the case of laboratory experiments where $\Delta m_{ij}^2 L/4E << 1$ for 
$i<j<4$ but $m_4^2 L/4E$ is not necessarily large or small, the corresponding 
formula is
\begin{equation}
P_{\alpha \beta} = \delta_{\alpha \beta} \left[ 1 - 2 U_{\alpha 4}^2 
\left( 1 - x \cos {m_4^2 L \over 2 E} \right) \right] + U_{\alpha 4}^2 
U_{\beta 4}^2 \left[ 1 - 2x \cos {m_4^2 L \over 2 E} + x^2 \right].
\end{equation}
Note that the above expression simplifies to a function of $U_{\alpha 4}$, 
$U_{\beta 4}$, and $x$ if $m_4$ is large, and to a function of $U_{\alpha 4}$ 
and $U_{\beta 4}$ alone if $x=0$ whatever the value of $m_4$.  In those 
circumstances, the corresponding laboratory experiment has no sensitivity to 
oscillations, but does measure one fixed number.  Specifically, if $m_4$ is 
large, then
\begin{equation}
P_{\mu e} = U^2_{e4} U^2_{\mu 4} (1+x^2), ~~~ P_{ee} = (1-U_{e4}^2)^2 + x^2 
U_{e4}^4, ~~~ P_{\mu \mu} = (1-U_{\mu 4}^2)^2 + x^2 U_{\mu 4}^4.
\end{equation}
If $x=0$, then regardless of $m_4$, Eq.~(8) reduces to Eq.~(9) but with $x$ 
set equal to zero.  The LSND experiment obtains\cite{3}
\begin{equation}
P_{\mu e} = 3.1 \begin{array} {l} + 1.1 \\ -1.0 \end{array} \pm 0.5 \times 
10^{-3},
\end{equation}
whereas BNL-E734 has\cite{10} $P_{\mu e} < 1.7 \times 10^{-3}$ and BNL-E776 
has\cite{11} $P_{\mu e} < 1.5 \times 10^{-3}$.  Using the LSND 90\% 
confidence-level limit of $P_{\mu e} > 1.3 \times 10^{-3}$, we find therefore 
reasonable consistency among these experiments.  [The most recent result of 
the ongoing KARMEN II experiment\cite{18} is $P_{\mu e} < 2.1 \times 10^{-3}$, 
which will eventually have the sensitivity to test Eq.~(10).]  The recent 
CCFR experiment\cite{12} measures $P_{\mu e} < 
0.9 \times 10^{-3}$, but its average $L/E$ is one to two orders of magnitude 
smaller than those of the other experiments, hence its $x$-value may be taken 
to be close to one and the usual oscillation interpretation of the data 
holds.  This constraint implies that $m_4^2 < 30$ eV$^2$.

At $m_4 \sim 5$ eV, we are below the CCFR exclusion and in a marginal region 
of the parameter space for pure neutrino oscillations consistent with the LSND 
evidence and the exclusion from BNL-E734 and BNL-E776.  Between $m_4 \sim 5$ 
eV and $m_4 \sim 3$ eV, the BNL-E734 data exclude a solution if $x = 1$ and 
because that experiment has an average $L/E$ an order of magnitude smaller 
than that of BNL-E776, LSND, or CDHSW, the decay factor goes against having 
a consistent solution here even if $x < 1$.  Below $m_4 \sim 3$ eV, the 
oscillation + decay interpretation of the latter 3 experiments becomes 
important, as shown below. 

Ideally, one should reanalyze the results of all the laboratory experiments 
using Eq.~(8) and verify whether the positive LSND signal can coexist 
with the exclusion limits from the other laboratory experiments by extending 
the usual parameter space of $m_4$, $U_{e4}$, and $U_{\mu 4}$ to include 
$\tau_4$ as well.  This can be done only by using the full data set of each 
of the experiments and is best performed by the experimenters themselves.  
In the absence of such a calculation, we point out here the crucial fact 
that the CDHSW experiment\cite{8} would see no difference in its two 
detectors at distances of 130 m and 885 m, if the effective values of the 
quantity $\exp (-m_4L/2E\tau_4) \cos (m_4^2 L/2 E)$ is the same.  In Table I, 
we show $\Gamma_4/m_4 (= 1/\tau_4 m_4)$ as a 
function of $m_4^2$ near 6 eV$^2$ for which this happens, using as our very 
crude approximation the fixed values of $L_1/E = 0.065$ m/MeV and $L_2/E = 
0.442$ m/MeV.  This illustrates the possibility that the decrease from $x_1$ 
to $x_2$ due to decay may be compensated by the increase in the value of the 
cosine from $L_1$ to $L_2$ due to oscillations.  Note also that there is a 
range of $m_4^2$ for which a null solution exists with varying $\Gamma_4/m_4$, 
whereas if the latter is zero, then $m_4^2$ has only discrete solutions 
(at 4.8 and 6.6 eV$^2$ for example).  In the realistic case of 
integrating over the experimental energy spectrum, both solutions will be 
smeared out, but the possibility of decay should result in a larger range of 
acceptable values of $m_4^2$.
For consistency, we also show in Table I the values of $f \equiv 
P_{\mu e}/U^2_{e4} U^2_{\mu 4} = 1 - 2x \cos (m_4^2 L /2 E) + x^2$ 
for the LSND and BNL-E776 experiments, using the fixed values of $L/E$ = 0.75 
and 0.5 m/MeV respectively.  This shows that the value of $P_{\mu e}$ as seen 
by the LSND experiment can be larger than that of BNL-E776 for $4.8 < m_4^2 < 
5.8 $ eV$^2$.
  
To discuss solar and atmospheric neutrino oscillations, let us focus on the 
following specific model.  Let $\cos \theta = \sqrt {2/3}$ and $\sin \theta 
= \sqrt {1/3}$ in Eqs.~(1) and (2), and let $\nu_s$ mix with $\nu_2$ only, 
then $U_{\alpha i}$ is given by
\begin{equation}
U = \left[ \begin{array} {c@{\quad}c@{\quad}c@{\quad}c} \sqrt {2/3} & 
c \sqrt {1/3} & 0 & s \sqrt {1/3} \\ - \sqrt {1/6} & c \sqrt {1/3} & 
- \sqrt {1/2} & s \sqrt {1/3} \\ - \sqrt {1/6} & c \sqrt {1/3} & \sqrt {1/2} 
& s \sqrt {1/3} \\ 0 & -s & 0 & c \end{array} \right],
\end{equation}
where $c$ and $s$ are respectively the cosine and sine of the $\nu_s - \nu_2$ 
mixing angle.  For solar neutrino oscillations, we have
\begin{equation}
P_{ee} = \left( 1 - {s^2 \over 3} \right)^2 - {4 \over 9} (1-s^2) \left( 1 - 
\cos {\Delta m^2_{12} L \over 2 E} \right) + {x^2 s^4 \over 9}.
\end{equation}
In the limit $s = 0$, this reduces to the usual two-neutrino formula with 
$\sin^2 2 \theta = 8/9$ which is a good fit to the data\cite{2}, either as 
the large-angle matter-enhanced solution or the vacuum oscillation solution. 
With a small $s^2/3$ of order a few percent [between 0.026 ($x=1$) and 0.037 
($x=0$) for $P_{\mu e}$(LSND) = $1.35 \times 10^{-3}$], this is definitely 
still allowed.  Note that this result is not sensitive at all to the last term 
because $s^4/9$ is of order $10^{-3}$.

For atmospheric neutrino oscillations, we have
\begin{equation}
P_{ee} = \left( 1 - {s^2 \over 3} \right)^2 + {x^2 s^4 \over 9}, ~~~ 
P_{e \mu} = P_{\mu e} = (1 + x^2) {s^4 \over 9},
\end{equation}
\begin{equation}
P_{\mu \mu} = \left( 1 - {s^2 \over 3} \right)^2 - {1 \over 2} \left( 1 - 
{2 s^2 \over 3} \right) \left( 1 - \cos {\Delta m^2_{23} L \over 2 E} \right) 
+ {x^2 s^4 \over 9}.
\end{equation}
Here the limit $s = 0$ corresponds to the canonical $\nu_\mu \to \nu_\tau$ 
solution with $\sin^2 2 \theta = 1$.  As it is, the prediction of $\nu_e \to 
\nu_e$ is still a fixed number, but smaller than unity (0.93 for $s^2/3 = 
0.037$).  Given that there is an uncertainty of about 20\% in the absolute 
flux normalization, we should consider instead the ratio
\begin{equation}
{2 P_{\mu \mu} + P_{e \mu} \over P_{ee} + 2 P_{\mu e}} \simeq 2 \left[ 1 - 
{s^4 \over 6} - {1 \over 2} \left( 1 - {2 s^4 \over 9} \right) \left( 1 - 
\cos {\Delta m^2_{23} L \over 2 E} \right) \right],
\end{equation}
where we have made an expansion in powers of $s^2$ and assumed that the ratio 
of $\nu_\mu$ to $\nu_e$ produced in the atmosphere is two.  It is clear that 
this is numerically indistinguishable from the case $s = 0$ .
 
In this model, the decay $\nu_4 \to \bar \nu_2 + M$ has some very interesting 
experimental consequences.  For example, $\nu_e$ from the sun decays through 
its $\nu_4$ component into $\bar \nu_2 = (c/\sqrt 3) (\bar \nu_e + 
\bar \nu_\mu + \bar \nu_\tau) - s \bar \nu_s$.  Hence
\begin{equation}
P(\nu_e \to \bar \nu_e) = P(\nu_e \to \bar \nu_\mu) = P(\nu_e \to 
\bar \nu_\tau) = {s^2 c^2 \over 9} \sim 10^{-2},
\end{equation}
where the energy of $\bar \nu_\alpha$ is only 1/2 that of $\nu_e$ and $x=0$ 
has been assumed.  This 
is in principle detectable especially since the $\bar \nu_e p$ capture cross 
section is about 100 times that of $\nu_e e$ scattering at a few MeV.  
Unfortunately, the Super-Kamiokande experiment has an energy threshold of 6.5 
MeV for the recoil electron and taking into account the additional 1.8 MeV 
threshold for the $\bar \nu_e p \to e^+ n$ reaction, this would require the 
original $\nu_e$ energy to be above 16.6 MeV, placing it outside the solar 
neutrino spectrum.  With the recently lowered Super-Kamiokande energy 
threshold of 5.5 MeV, the fraction of solar $\nu_e$ above 14.6 MeV is $1.6 
\times 10^{-4}$.  Given the small probability of $P(\nu_e \to \bar \nu_e)$, 
this will not change appreciably the total number of observed $e$-like events. 
Regardless of energy threshold, the inability of Super-Kamiokande to 
distinguish $e^+$ from $e^-$ or to detect the 2.2 MeV photon from neutron 
capture on free protons makes it difficult to pin down this possibility in 
any case. 

In the Sudbury (SNO) neutrino experiment\cite{19}, the energy threshold for 
detecting recoil electrons is 5 MeV, but since there is also a threshold 
of about 4 MeV for breaking up the deuterium nucleus into two neutrons and 
a positron, the neutrino energy required is more than about 18 MeV.  This 
again places it outside the solar neutrino spectrum.  On the other hand, if 
the experimental energy threshold can be significantly lowered, then SNO may 
be able to see this effect because the $\bar \nu_e$ signature ($\bar \nu_e + d 
\to n + n + e^+$) is distinct from that of $\nu_e$. 

The best chance for detecting antineutrinos from the decay of $\nu_4$ is 
offered by the BOREXINO experiment\cite{20} with a very low energy threshold 
of 0.25 MeV.  Taking into account the 1.8 MeV needed for inverse beta decay, 
i.e. $\bar \nu_e p \to e^+ n$, this means that solar neutrinos with energy 
above 4.1 MeV can be detected as antineutrinos.  
The idea of looking for antineutrinos from the sun was motivated by the 
possibility of a large neutrino magnetic moment which may convert $\nu_e$ 
into $\bar \nu_e$ in the sun's magnetic field.  The capability 
of BOREXINO for detecting this has been discussed earlier\cite{21}.  For 
our new distinctive effect of $\nu_4$ decay, the observed antineutrino energy 
spectrum is predicted to go from $f(E)$ to $f(E/2)$, where $E$ is the energy 
of the original neutrino.

For atmospheric neutrinos, since $\bar \nu_\mu$ and $\bar \nu_e$ are produced 
together with $\nu_\mu$ and $\nu_e$ in about equal amounts, it is not 
possible to tell if a given event comes from the primary neutrino or its 
decay product, even if the detector could measure the charge of the 
observed lepton.

To search for the $\nu_\mu \to \bar \nu_e$ transition in the LSND and 
KARMEN experiments, one would use the monoenergetic (29.8 MeV) $\nu_\mu$ 
from $\pi^+$ decay at rest, which has the signature of a monoenergetic 
positron of 13.1 MeV from inverse beta decay, i.e. $\bar \nu_e p \to e^+ n$, 
in coincidence with a 2.2 MeV photon from the subsequent capture of the 
neutron by a free proton.  However, this signal is overwhelmed by the 
neutral-current reaction $\nu ~^{12}C \to \nu ~ ^{12}C^*$, with the subsequent 
emission of a  15.1 MeV photon.

In proposed long-baseline $\nu_\mu \to \nu_\tau$ appearance experiments, 
the oscillation probability is given by
\begin{equation}
P_{\mu \tau} = \left( 1 - {s^2 \over 3} \right)^2 - {1 \over 2} \left( 1 - 
{2 s^2 \over 3} \right) \left( 1 + \cos {\Delta m^2_{23} L \over 2 E} \right) 
+ {x^2 s^4 \over 9},
\end{equation}
which is not easily distinguished from the $s = 0$ case.  However, the decay 
products of $\nu_4$, i.e. $\bar \nu_e$, $\bar \nu_\mu$, and $\bar \nu_\tau$, 
may be observable with their own unique signatures, depending on the 
capabilities of the proposed detectors.

In the case of four-neutrino oscillations, the effective number of neutrinos 
$N_\nu$ in Big Bang Nucleosynthesis is an important constraint\cite{22}.  In 
this model, with $m_4 \sim$ few eV and $s^2 \sim$ few percent, the presence of 
a stable $\nu_s$ would have counted as an extra neutrino species, making 
$N_\nu = 4$. 
This may not be acceptable if $N_\nu < 4$, as indicated from the observed 
primordial $^4$He abundance\cite{23}.  The decay of $\nu_4$ changes $N_\nu$ 
to 3 + the contribution of the Majoron (i.e. 4/7).

With $\nu_4$ as a component of $\nu_e$, neutrinoless double decay has an 
effective $\nu_e$ mass of $(s^2/3)m_4 \sim 0.2$ eV if $m_4 \sim 5$ eV. 
This value is just at the edge of the most recent experimental upper 
bound\cite{24}.

Finally a comment on the neutrino contribution to dark matter may be in order. 
With $\nu_4$ decaying and $m_1$, $m_2$, and $m_3$ being too small, there is 
no neutrino dark matter.  However, it is possible that $m_1 \simeq m_2 \simeq 
m_3 \simeq$ few eV, while $m_4$ is higher by another few eV, in which case 
$\nu_1$, $\nu_2$, and $\nu_3$ will contribute to dark matter.  Our discussion 
goes through almost unchanged, except that $m_4^2$ in Eq.~(8) will be 
replaced by $m_4^2 - m_{1,2,3}^2$.

In conclusion, we have shown in this paper that a hierarchical four-neutrino 
scenario is acceptable as a solution to all present neutrino data regarding 
the disappearance and appearance of $\nu_e$ and $\nu_\mu$.  The assumed 
singlet neutrino of a few eV may decay into a linear combination of the three 
known doublet neutrinos with half of the energy.  This new feature allows 
our proposal to be tested in future solar neutrino experiments such as 
BOREXINO (and perhaps SNO), and should be considered in forthcoming 
long-baseline accelerator neutrino experiments.

\vspace{0.3in}
\begin{center}{ACKNOWLEDGEMENT}
\end{center}

One of us (G.R.) thanks the Physics Department, University of California, 
Riverside, for hospitality while this work was done.  The research of E.M. 
was supported in part by the U.~S.~Department of Energy under Grant 
No.~DE-FG03-94ER40837.

\newpage
\bibliographystyle{unsrt}

\newpage

\begin{table}
\begin{center}
\begin{tabular}{|c|c|c|c|}
\hline
$m_4^2$(eV$^2$) & $\Gamma_4/m_4$ & $f$(LSND) & $f$(E776) \\
\hline
4.8 & 0 & 3.92 & 0.04 \\
\hline
5.0 & 0.030 & 3.04 & 0.04 \\
\hline
5.2 & 0.065 & 2.21 & 0.19 \\
\hline
5.4 & 0.085 & 1.72 & 0.38 \\
\hline
5.6 & 0.095 & 1.37 & 0.57 \\
\hline
5.8 & 0.095 & 1.09 & 0.78 \\
\hline
6.0 & 0.086 & 0.54 & 1.03 \\
\hline
6.2 & 0.068 & 0.55 & 1.37 \\
\hline
6.4 & 0.038 & 0.22 & 1.94 \\
\hline
6.6 & 0 & 0.0 & 3.0 \\
\hline
\end{tabular}
\vskip 0.5in
\caption{Null solution for oscillation and decay at the two CDHSW detector 
distances.}
\end{center}
\end{table}

\end{document}